\pdfoutput=1
\RequirePackage{fix-cm}
\documentclass[smallextended]{svjour3}       
\smartqed  
\usepackage{graphicx}
 \journalname{EXP ASTRON}

 \usepackage{epstopdf}
\usepackage{epsfig}          
\usepackage[numbers]{natbib}         
\usepackage{amssymb}        
\usepackage{color}           
\usepackage{url}             
\usepackage[T1]{fontenc}        
\usepackage{geometry}
\usepackage{hvfloat}
\usepackage{graphicx}

\begin{document}

\title{The soft X-ray spectrometer polarimeter SolpeX}
\titlerunning{The soft X-ray spectrometer polarimeter SolpeX}


\author{J.~Sylwester \and
        M.~{St{\c e}{\'s}licki} \and
        J.~B\k{a}ka{\l}a \and
        S.~P{\l}ocieniak \and
        \.{Z}.~Szaforz \and
        M.~Kowali\'{n}ski \and
        D.~{\'S}cis{\l}owski \and
        P.~Podg\'{o}rski \and
        T.~Mrozek \and
        J.~Barylak\and
        A.~Makowski\and
        M.~Siarkowski \and
        Z.~Kordylewski \and
        B.~Sylwester \and
        S.~Kuzin \and
        A.~Kirichenko \and
        A.~Pertsov \and
        S.~Bogachev
}

\authorrunning{J.~Sylwester {\it et al.}}

\institute{J.~Sylwester  \and M.~{St{\c e}{\'s}licki} \and J.~B\k{a}ka{\l}a \and S.~P{\l}ocieniak \and  \.{Z}.~Szaforz \and M.~Kowali\'{n}ski \and D.~{\'S}cis{\l}owski \and   P.~Podg\'{o}rski \and T.~Mrozek \and J.~Barylak \and A.~Makowski \and  M.~Siarkowski \and Z.~Kordylewski \and  B.~Sylwester
 \at Space Research Centre, Polish Academy of Sciences\\ Bartycka 18A, 00-716 Warsaw, Poland\\
  \email{js@cbk.pan.wroc.pl}    \newline\newline
   S.~Kuzin \and A.~Kirichenko \and  A.~Pertsov \and S.~Bogachev \at P.N. Lebedev Physical Institute, Russian Academy of Sciences\\ Leninsky Prospekt 53, Moscow 119991, Russia\\
  \email{kuzin@lebedev.ru}\\
                                   }

\date{Received: date / Accepted: date}

\maketitle

\begin{abstract}
We present a novel X-ray assembly of functionally related instrument blocks intended to measure solar flare and active region (AR) spectra from within the Russian instrument complex KORTES, to be mounted aboard the {\em International Space Station} ({\em ISS}). SolpeX consists of three blocks: fast-rotating multiple flat crystal Bragg spectrometer, pin-hole X-ray spectral imager and Bragg polarimeter. This combination of measuring blocks will offer an opportunity to detect/measure possible X-ray polarization in soft X-ray emission lines/continuum and record spectra of solar flares, in particular during their impulsive phases. Polarized Bremsstrahlung and line emission may arise from presence of directed particle beams colliding with denser regions of flares. As a result of evaporation, the X-ray spectral-components are expected to be Doppler shifted, which will also be measured.

In this paper, we present details of the construction of three SolpeX blocks and discuss their functionality.
Delivery of KORTES with SolpeX to {\em ISS} is expected in 2020/2021.
\end{abstract}


\section{Introduction}
The conversion (due to reconnection) of accumulated magnetic energy in solar active regions often leads to formation of beams of accelerated particles. Interaction of these beams with denser plasmas (target) leads to the formation of transient sources with spectra extending over the entire electromagnetic range from radio to gamma rays. Studies of X-ray flare emission was and is the most direct way to infer properties of these particle beams and thermal hot, multimillion degree plasmas.
Studying the nature of flare X-ray sources is crucial for understanding the energy release and its conversion to thermalised, turbulent and kinetic components. Accelerated particle beams, while interacting with ambient denser plasma, release a small amount of energy in the form of non-thermal Bremsstrahlung, expected to be polarized due to anisotropic character of interactions. With SolpeX, we intend to detect the polarization for the first time in the soft X-ray lines and continuum emission at $\sim$3\,keV, which will additionally constrain the properties of electron beams by determining the orientation of the magnetic field at the interaction region \cite{Emslie_etal2008}.
Models of interaction predict that the harder X-ray emission, consisting solely of continuum emission, should be highly polarized, with a polarization degree as high as 40\% at 20\,keV \cite{Zharkova_etal2010}. The soft X-ray emission (at energies $<10$\,keV) consists of lines as well as continuum, and the line emission is also expected to be polarized \cite{Elvert_Haug1970}. The X-rays from anisotropic interactions are expected to dominate during the early impulsive phase of flares (the first seconds to minutes of flares). The other processes contributing to polarization, although at a lower level, and the later flare phases are due to anisotropy of the electron distribution function within thermalized sources \cite{Emslie_Brown1980} and/or backscattering of the X-rays formed in the coronal X-ray source on lower-lying structures of solar atmosphere \cite{Jeffrey_Kontar2011}.

The importance of measuring polarization of X-ray flare emission was realized from the very beginning of the space era. The first attempts at detecting X-ray polarization from solar flares were made using polarimeters on-board Soviet {\em Intercosmos} satellites. The early results indicated rather high values of average polarization $P=40 (\pm 20)\%$ \cite{Tindo_etal1970} and $20\%$ \cite{Tindo_etal1972a}, \cite{Tindo_etal1972b} at energies $\sim$15\,keV. Subsequent observations with an X-ray polarimeter on board the {\em OSO-7} satellite point to smaller value of the polarization~10\% \cite{Nakada_etal1974}. Later studies show even smaller values of polarization often below sensitivity threshold. The polarimeter on-board {\em Intercosmos--11} \cite{Tindo_etal1976} detected polarization of few percent only at the energy $\sim$15\,keV. Similar values of the degree of polarization were obtained from the polarimeter placed on-board {\em STS-3} mission. The upper limits determined in the 5--20\,keV energy range by \cite{Tramiel_etal1984} were 2.5\% to 12.7\%.

The most recent determinations of solar flare X-ray polarization at the higher energy range were obtained from analysis of data collected by {\em RHESSI} (NASA) and the SPR-N polarimeter on-board the {\em CORONAS-F} satellite.

Using {\em RHESSI} data \cite{McConnell_etal2003} found a polarization of  $\sim$18$\%$ in the energy range 20--40\,keV. The source was an X-class solar flare on July 23, 2002 at 00:35~UT (SOL2002-07-23T00:35). For the same event and an additional X-class flare (SOL2003-10-28T11:06) the polarization at levels of $21\% \pm 9\%$ and $11\% \pm 5\%$ were reported by \cite{Boggs_etal2006} at much higher energies (0.2--1\,MeV).
\cite{Suarez-Garcia_etal2006} found for six X-class flares and one M-class flare values for the polarization degree in the range between 2\% and 54\% at an energy range between 100 and 350\,keV.

According to \cite{Bogomolov_etal2004} the SPR-N polarimeter indicated strong (>70\%) polarization at energies 40--100~keV for a flare on 29 October 2003, but zero polarization for two other flares (SOL2013-10-28T11:00 and SOL2013-11-04T19:34). Using data from this same satellite Zhitnik et al. \cite{Zhitnik_etal2006} found, among 90 analyzed flares, one event (X10 class flare SOL2003-10-29T20:37) which showed a significant polarization degree exceeding 70\% at energies 40--100\,keV and about 50\% at lower energies (20--40\,keV). For other 25 events, the upper limits for the polarization were estimated to be between 8\% and 40\%.

All these X-ray polarization determinations were obtained using a Compton scattering technique. For polarized radiation, the azimuthal distribution of the scattered photons is no longer isotropic, but is related to the polarization vector of the incident photons. The energy resolution of the Compton polarimeters mentioned was rather coarse. Also, a low signal-to-noise ratio did not allow high-time-resolution observations.

\begin{figure}
\centerline{\includegraphics[,width=1.0\textwidth,clip=]{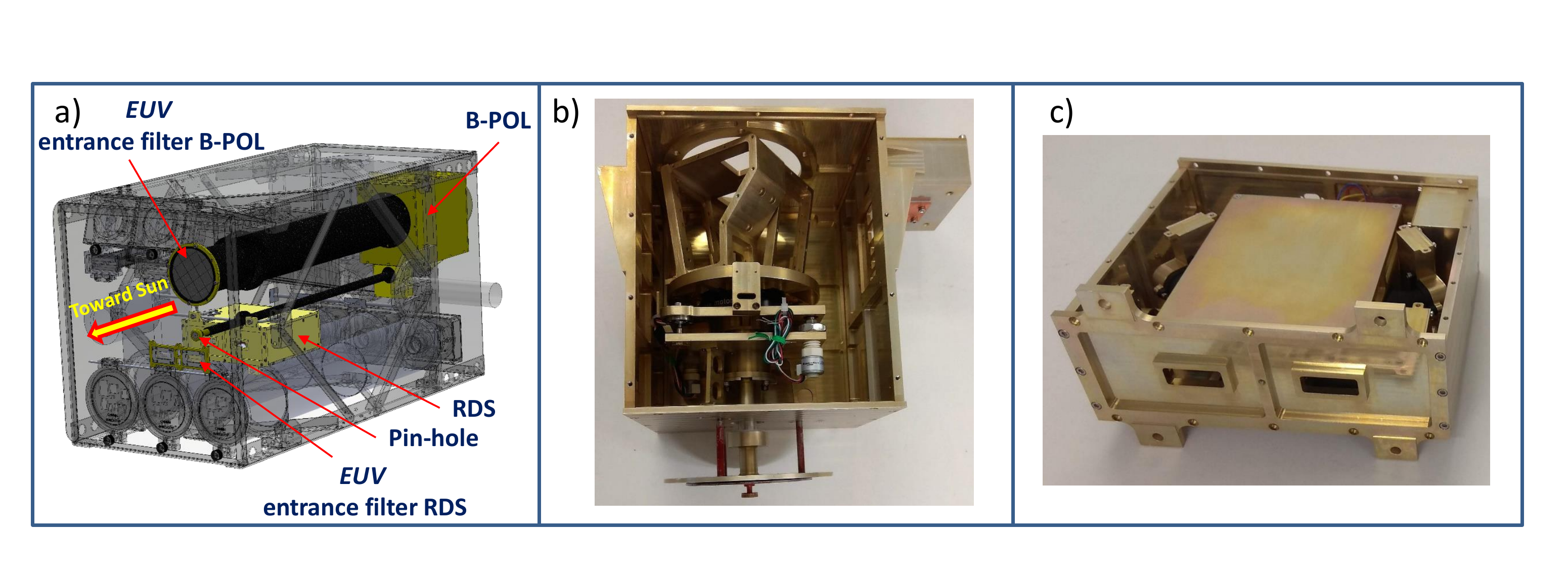}}
\caption{{\bf a)} Location of pin-hole, B-POL (Bragg-POLarimeter) and RDS (Rotating Drum Spectrometer) blocks within the Russian-build KORTES instrument;
{\bf b)} The engineering model of B-POL polarimeter;
{\bf c)} The engineering model of RDS.}
\label{fig_kortes}
\end{figure}

In view of rather equivocal results obtained so far, we devised a different technique of polarization detection with the aim of studying the polarization in a softer X-ray band, which relies on the selective properties of Bragg reflection (close to the Brewster angle), on the orientation of the polarization vector relative to the crystal plane (see Fig.~\ref{Bpol_inside}).

In the following, we describe in detail the construction of the three blocks comprising SolpeX, designed to observe solar soft X-ray emission i.e.:
\vspace{1mm}
\begin{itemize}
\item PHI -- a simple {\bf P}in-{\bf H}ole soft X-ray {\bf I}mager-spectrophotometer using detector with moderate spatial ($\sim$20\,arcsec), spectral (FWHM of Fe line at 6.4\,keV is 0.44\,keV, 0.37\,keV for Cu line at 8.05\,keV and 0.65\,keV for Mo line at 17.48\,keV) and high time resolution (1/8\,s)
\item B-POL the 3.9\,\AA--4.15\,\AA~spectrometer and polarimeter ({\bf B}ragg {\bf POL}arimeter) with a low 1-2\% linear polarization detection threshold.
\item RDS -- a fast {\bf R}otating {\bf D}rum X-ray {\bf S}pectrometer with high time resolution (0.1\,s)
\vspace{.5mm}
\end{itemize}

In Fig.~\ref{fig_kortes}a, we show the location of the units within the KORTES instrument (as of February 2018). The RDS unit is placed close to the radiator, which is used by the entire instrument. The B-POL polarimeter is placed in the rear section of KORTES. This location is imposed by functionality, as the polarimeter's CMOS detectors will be cooled down to below -20$^{\circ}$C using the rotating radiator attached to the rotating axis (which will also act as a heat sink). On the side of the polarimeter, a square CMOS detector is attached on which the X-ray image of the Sun from the pin-hole is projected. The pin-hole itself is mounted on the front-plane of KORTES. The engineering models of B-POL and RDS are shown in Fig.~\ref{fig_kortes}b and Fig.~\ref{fig_kortes}c respectively.

\begin{figure}
\centerline{\includegraphics[,width=1.0\textwidth,clip=]{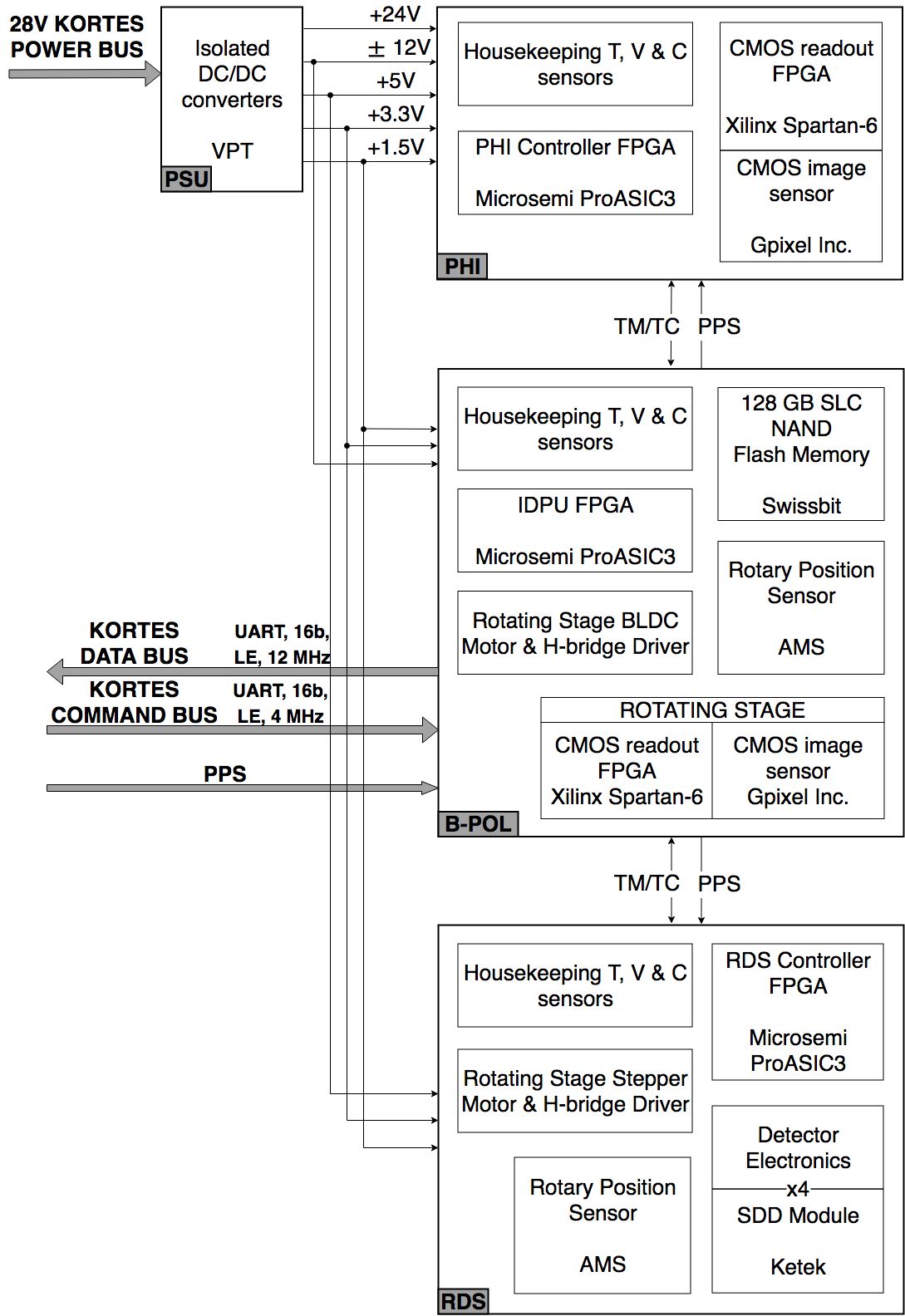}}
\caption{Block diagram of SolpeX electronics and connections to KORTES.}
\label{fig_solpex_blocks}
\end{figure}

A general diagram of the SolpeX instrument electronics indicating the main components is shown in Fig.~\ref{fig_solpex_blocks}. The B-POL block is treated as the master unit and has Telemetry/Telecommand (TM/TC) serial interfaces for direct connection to KORTES. In addition, the Pulse Per Second (PPS) signal from KORTES is distributed to the other SolpeX modules. Each of them synchronizes its internal, precise clocks with this signal, thus maintaining mutual time synchronization without the need to correct the clocks during operation. In order to perform these control functions of the instrument and to handle the incoming data, the polarimeter is equipped with a dedicated FPGA ProASIC3 chip as well as 128\,GB of Flash memory to buffer the data flow. Also noteworthy is the fact that the detector part of the polarimeter electronics is located on the rotating part of the device. For this reason, we have used a slip ring to provide an electrical connection for the power supply and data transmission for this part of the instrument. Details of the other components of the electronic blocks are described in the following sections.

From the electronics perspective, we also distinguish an additional block, the Power Supply Unit (PSU), common for the three SolpeX blocks. Based on VPT's series of high reliability DC/DC converters, it will produce the power required for the measurement blocks from the unregulated 28\,V KORTES power bus.

KORTES itself is to be placed on the solar pointing platform attached to the Russian {\em Nauka} module on {\em ISS}. The pointing platform of {\em Nauka} is guided towards the center of the solar disk. The illumination conditions at the placement of KORTES are not ideal for making solar observations as the complicated mechanical structure of {\em ISS} causes substantial vignetting. Only 10--12\,min of uninterrupted illumination is available for every {\em ISS} orbit, so catching the impulsive phase of a flare will need some luck. Estimates indicate that by 2021 the current solar minimum will have ended and that we might expect to observe the impulsive phases of  $\sim$40 C-class flares, 5~M-class flares and a single X-class flare during the first year of instrument operations. The KORTES instrument will be delivered to {\em ISS} by a cargo ship and then fixed to the \emph{Nauka} module by cosmonauts. After the mission is completed, the instrument will return to the ground for post-flight checks. The data from the instrument will be transmitted {\em on-line} to the recorder inside {\em ISS} using KORTES telemetry, so practically unlimited storage is available on hard disk. The solar X-ray image formed on CMOS will be displayed for the cosmonauts on board to follow solar activity in real time. A selected part of the information will be beamed down in real time, allowing prompt analysis leading to flight software corrections and upgrades. The instrument can be operated round the clock, also during the night-portions of the orbit for the calibration purposes or measurements of non-solar astrophysical sources.
Below we describe the three blocks of SolpeX in more detail.


\section{Pin-Hole soft X-ray Imager-spectrophotometer --- PHI}
\label{section:phi}
\begin{figure}
\begin{minipage}[b]{0.3\textwidth}
  \centering
  \includegraphics[width=\textwidth]{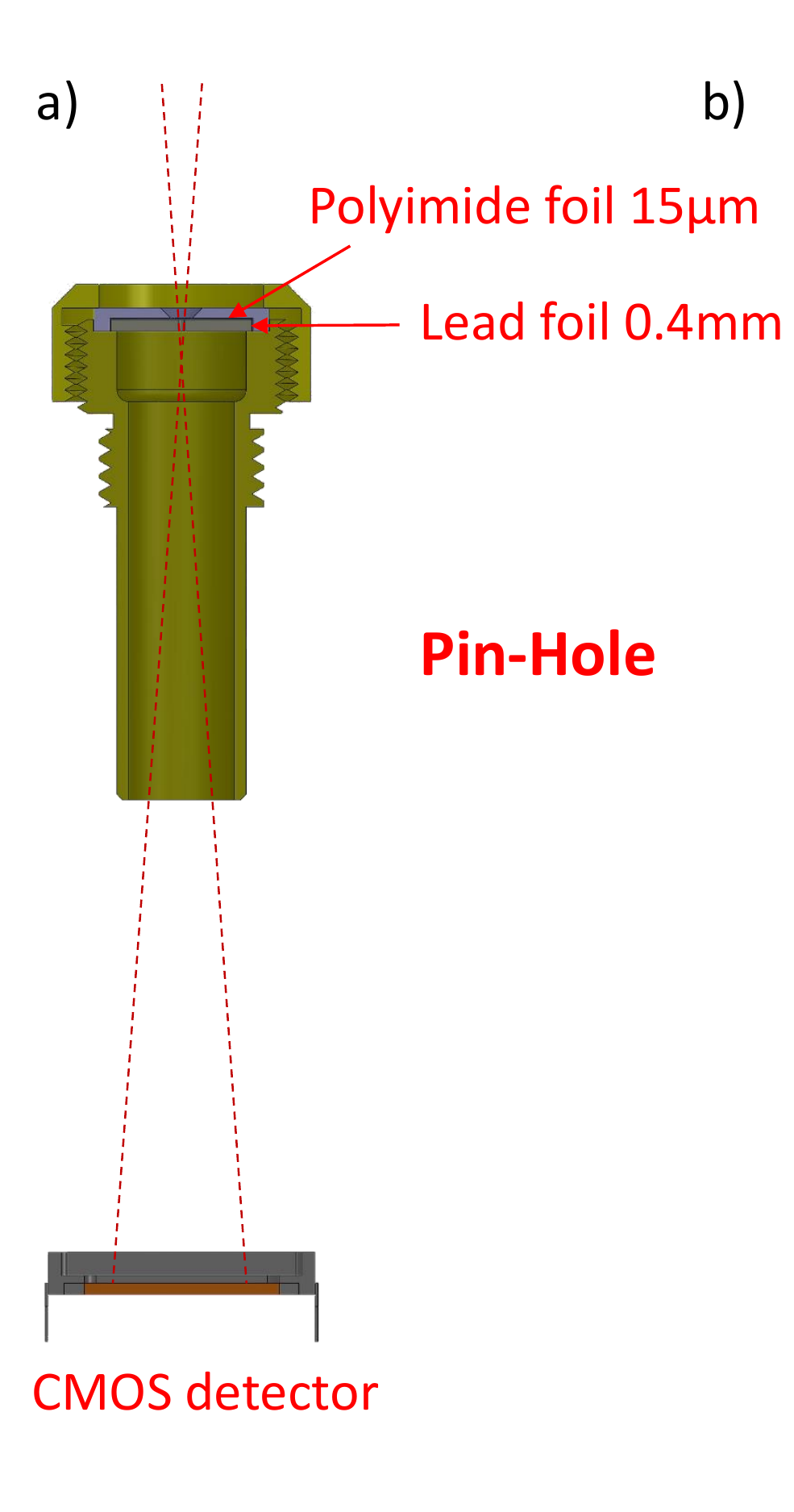}
\end{minipage}
 \hfill
\begin{minipage}[b]{0.7\textwidth}
  \centering
  \includegraphics[width=\textwidth]{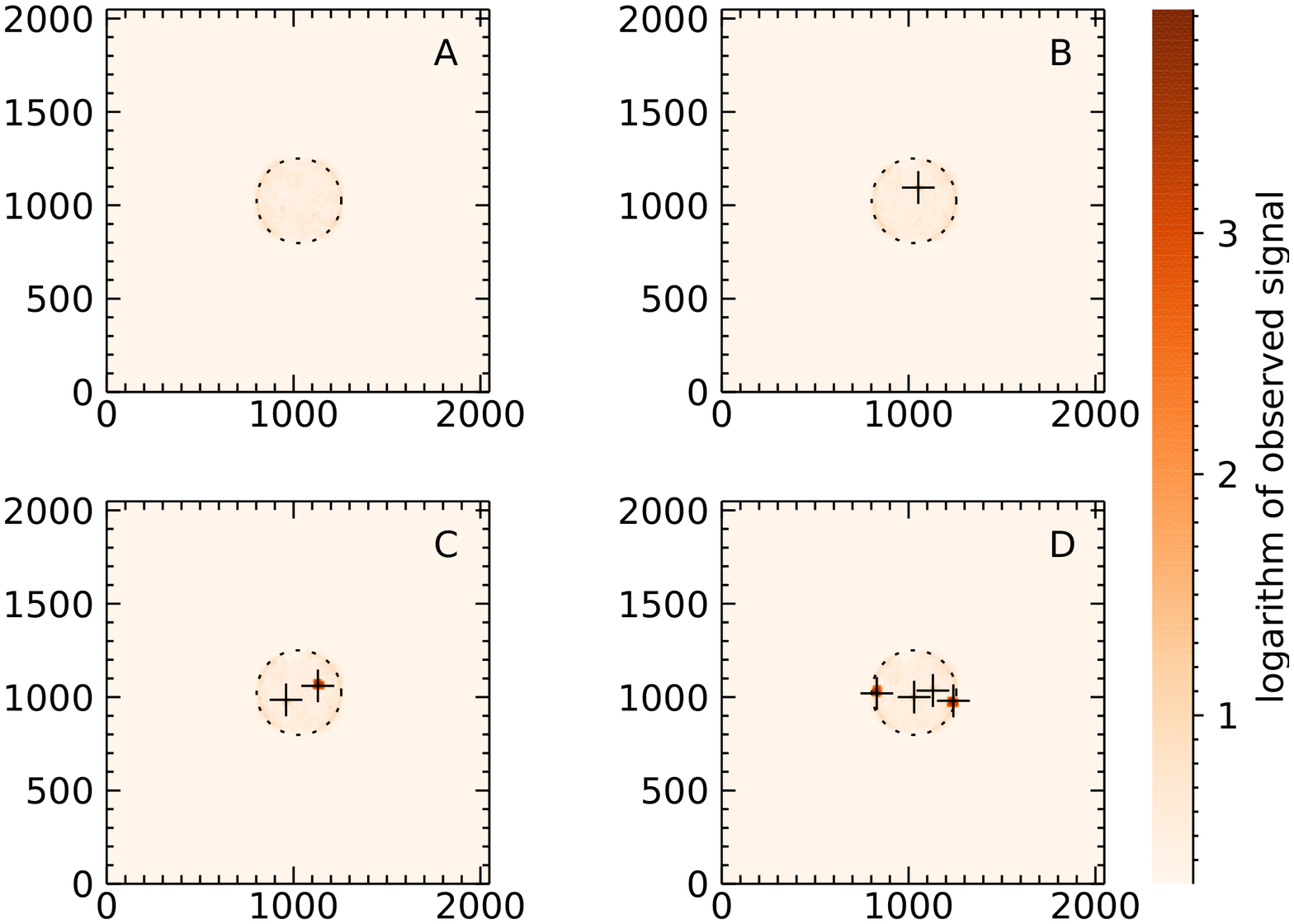}
\end{minipage}
\caption{{\bf a)} Photographs of the pin-hole - a cylindrical opening $\varphi$=0.7\,mm in the lead foil of 0.4\,mm thickness. This diameter of the pin-hole is adequate to achieve sufficient X-ray photon statistics. The pin-hole is covered by $\sim$15\,$\mu$m thick thermal filter as seen from the Sun direction. The X-ray image is projected on the CMOS. With the pin-hole - CMOS distance of 578\,mm, the projected diameter of the solar disc corresponds to $\sim$215~detector pixels.
{\bf b)} Simulated images of X-ray Sun for four characteristic situations: {\bf A} no active regions (AR) visible, {\bf B} an AR present, {\bf C} flare and AR present on separate locations on the disk and {\bf D} two flaring ARs with two M5 flares in progress on the opposite sites of the solar disc. The images shown are XRT Ti-poly coronal observations projected on the pin-hole camera CMOS. The on-board processing by the PHI electronics ``finds'' location of the solar limb (from limb brightening ring), position of the solar center in the CMOS coordinate system, determine location, sizes and intensities of ARs, as well as positions of flares and their intensities. }
\label{fig_pin-hole}
\end{figure}

Proper operation of the RDS and B-POL requires timely information on any solar X-ray activity that might be occurring on the solar disc. Such information can be obtained from a simple pin-hole telescope/imager/spectrometer equipped with thermal filter transmitting solar soft X-rays and rejecting solar thermal/optical/EUV emission. Such a filter made of aluminized polyimide will transmit in the range E>0.35\,keV, a similar spectral range to this covered by XRT Ti-poly filter on {\em Hinode} \cite{Golub_etal2007}. The image of the X-ray solar disk is formed behind the pin-hole on the square 2k$\times$2k CMOS (Gpixel Inc.). The readout time of the full frame can be between 1/48\,s and 1/8\,s. Most of the time the PHI will operate with the time resolution of 1/8\,s.

In Fig.~\ref{fig_pin-hole}b simulated X-ray images are shown as will be projected on the CMOS for four activity scenarios: no active regions visible, single AR present, flare and AR present on separate locations on the disk and two ARs with two flares in progress. This images were obtained by processing real XRT Ti-poly images. It is seen that limb brightening is clearly present in every case. The shape of the limb-brightened crescents allow (after on-board processing) for the determination of the disk edge and its centre in the CMOS coordinate system (see the limb position plotted as a circle in Fig.~\ref{fig_pin-hole}b). This important information is necessary to allow co-alignment with other images obtained by KORTES and can be used to analyse the accuracy of the fine solar pointing system of the {\em Nauka} module.

\begin{figure}
\centerline{\includegraphics[,width=1.0\textwidth,clip=]{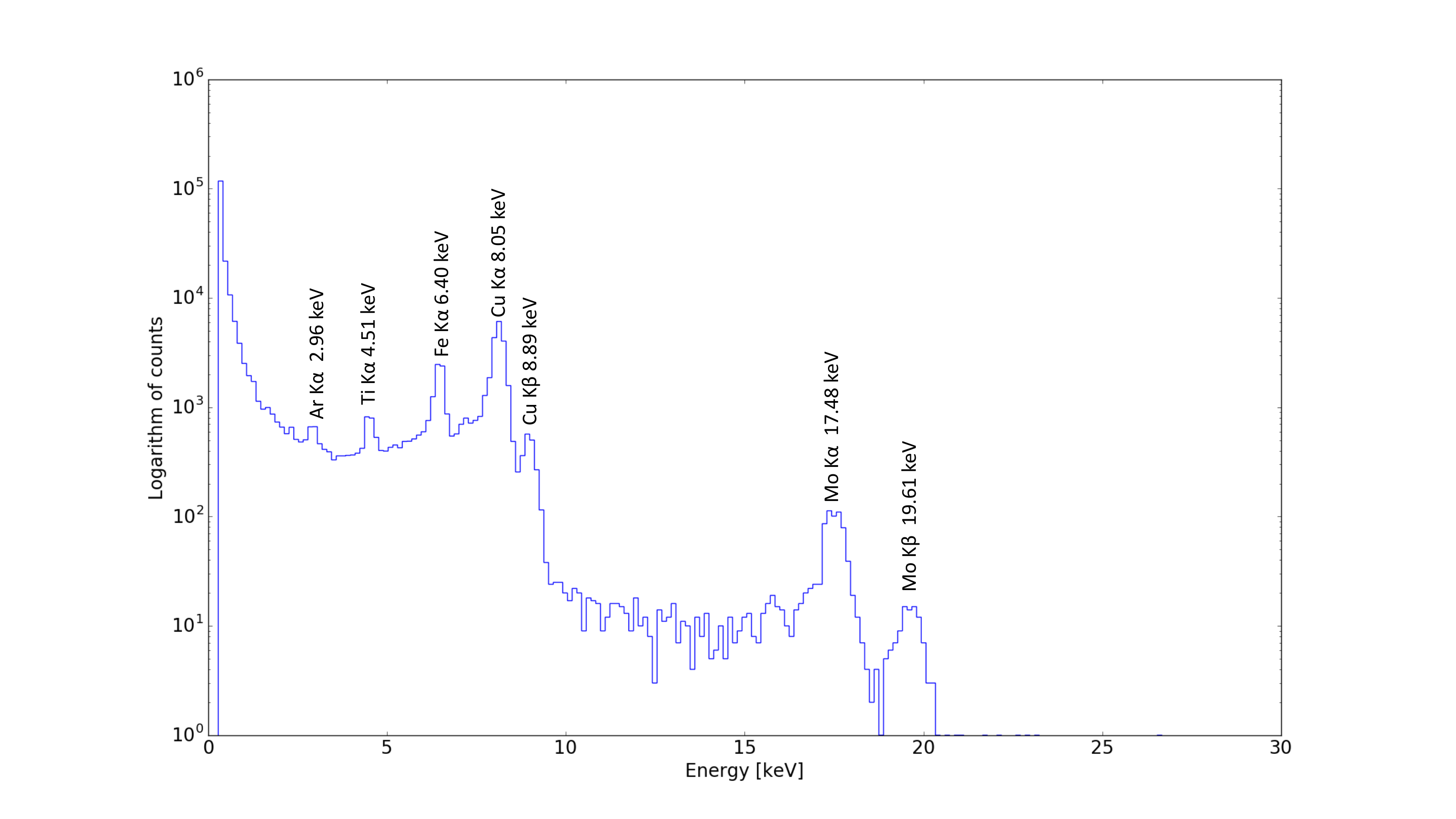}}
\caption{Histogram of CMOS energy spectrum taken in laboratory conditions. The Gpixel Inc. CMOS (GSENSE400 BSI) recorded the X-rays reflected from multi-metal foil system consisted of Ti, Fe, Cu and Mo. The foils were illuminated by the X-ray lamp set at 35\,keV voltage. Individual peaks due to characteristic K$\alpha$ emission are easily discernible. The ambient temperature during measurements was 20$^{\circ}$C. To avoid pileup the measurements were taken 120 times with 8\,ms of single exposition. The CMOS matrix was shielded from the direct illumination coming from the X-ray lamp itself. A readout and control system for this CMOS readout was based on FPGA Spartan 6 XCFSLX9. The CMOS readout time (see the text) allowed for a single photon/pixel regime. Such a mode will be routine for B-POL and most of PHI observing time except for the strongest flares (X-class).}
\label{fig_cmos}
\end{figure}

Using appropriate algorithms, actual positions of ARs and their sizes and orientation (from 2D Gaussian fits) will be determined as well as the total X-ray base-projected brightness for every resolved AR. The dispersion direction of the RDS crystal is aligned with the CMOS base, therefore the X-ray brightness shape along the dispersion will be recorded frequently and subsequently used to deconvolve the real spectral shapes from those observed by RDS (see Chapter \ref{section:rds}).

The CMOS to be used in B-POL and PHI is characterized by a good energy resolution, provided single X-ray photons strike on the pixel. The energy resolution of CMOS pixel is illustrated in Fig. \ref{fig_cmos}. With the frame read-out time 1/8\,s, a single photon/pixel scenario will take place most of the time, except for the most powerful flares, when double strikes/pixel are expected. Amplitude analysis of the signal from individual CMOS pixels will be used to discern solar X-ray photons from those due to secondary fluorescence or energetic particles. The brightness profile can be created for selected energies, that are the representative for individual spectral bands.

The X-ray corona projected on to the CMOS extends to $\sim$2.5 solar radii in both directions. A large portion of CMOS, away from the solar disk will not be illuminated by solar X-rays, allowing the changes of the orbital background and possible secondary X-rays formed within KORTES to be studied. A part of secondary X-ray contribution arising in the instrument due to energetic particle environment will be removed by covering the CMOS with a very thin plastic filter. This filter will prevent possible internal EUV and optical leaks to interfere with the solar X-ray image projected by a pin-hole. The background signal from CMOS will be used to estimate particle interference for the identical CMOS used in B-POL to record the X-ray spectra.

From images similar to that shown in Fig.~\ref{fig_pin-hole}b (to be collected every 1/8\,s) the soft X-ray light curves of individual active regions will be analyzed in real-time by the on-board software in order to detect and locate ARs and flares on the disc. A 2D Gaussian elliptical profile will be fitted to individual stronger brightenings after collecting enough statistics and their central positions will be determined. Those positions will be passed to the KORTES and B-POL internal pointing system in order to lock the orientation of the polarimeter rotation axis on selected targets.

The data obtained from the PHI, in particular the X-ray intensities from the total CMOS, background corrected signal and on-board detected individual AR(s) fluxes, will be used to determine the state of the X-ray corona, i.e. corresponding fluxes in at least two energy bands. The signal-to-noise ratios, as obtained from images in these two bands, will allow the determination of the average coronal temperatures (using a filter ratio technique) and corresponding average emission measures (from the X-ray flux)\cite{Gburek2013}. These data will also enable fluxes in the standard {\em GOES} X-ray bands 0.5--4\,\AA~and 1--8\,\AA~to be imitated on-board.

The total level and distribution of the emission on the disk will be used to define and classify the status of solar momentary activity into three classes:
\begin{itemize}
\item ``Quiet'' - only limb brightening + bright points contribution are present;
\item AR1 - one non-flaring AR detected by the algorithm (AR2, AR3 etc. if more AR detected);
\item FL1 - one flare in progress (or FL2, as sometimes two flares may happen at the same time in different ARs). AR1 + FL1 would mean one non-flaring and separate flaring AR present.
\end{itemize}

The history (flux levels from five previous time periods) of total X-ray signal recorded for individual components (total solar fluxes and fluxes in individual ARs - if present) will be used to assign ``phase'' of active Sun. This will be done by an on-board computer algorithm analysing the time dependences from these five preceding time points. If the fluxes are rising fast enough, the flare start is fixed and the rise phase is flagged; if maximum is reached, the flare maximum time is fixed. For monotonically decreasing fluxes the flare decay is flagged. The algorithm makes allowance for possible statistical fluctuations of the signal and works for the total X-ray signal from the Sun (pair of fluxes in two energy bands) as well as for individual ARs. If a flare is detected in particular position, the algorithm identifies the location of the centroid of the flaring AR and passes this information to the B-POL (and other KORTES blocks) in order to change the operation modes and repoint the B-POL rotation axis to lock on the flaring region.


\section{The Rotating Drum Spectrometer}
\label{section:rds}
\begin{figure}
\centerline{\includegraphics[,width=1.0\textwidth,clip=]{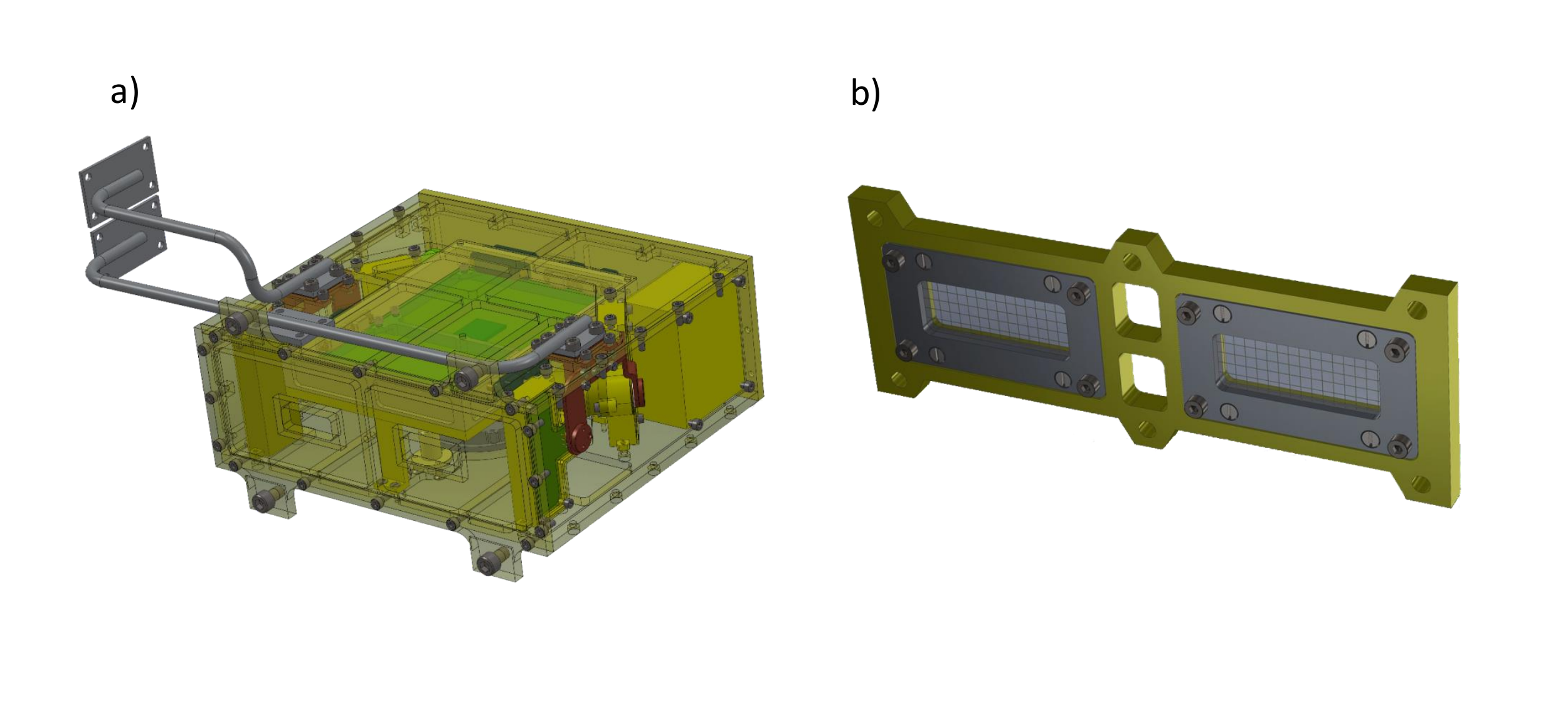}}
\caption{{\bf a)} RDS measurement block with the active cooling pipe system in grey {\bf b)} The RDS aperture filter consists of the two entrance windows. Filters limits UV, visible, and IR radiation from reaching the inside of KORTES, while not overly attenuating X-rays in the range from 0.4 to 23\,keV. The transmission of the filter in this spectral range is approximately 75\% or better. The filters are made of very thin polycarbonate (Lexan) 2000\,\AA~foils coated on both sides with 0.12\,$\mu$m (1200\,\AA) of aluminium. The polycarbonate window is supported by Kevlar mesh where the thickness of the thread is 30\,$\mu$m. Filter dimensions are 38$\times$15\,mm.}
\label{fig_rds}
\end{figure}

The fast-{\bf R}otating {\bf D}rum X-ray {\bf S}pectrometer (RDS) will allow us to investigate possible very fast changes in solar spectra during impulsive phases of flares. The RDS is located close to the front panel of KORTES with its detectors connected to the radiators with efficient active cooling pipes (see Fig.~\ref{fig_rds}a). It will allow the creation of an atlas of contiguous solar spectra in the entire soft X-ray range between 0.4\,\AA~and 23\,\AA~for various plasma temperatures. Based on such spectral atlas, it will be possible to study distribution of differential emission measure for various plasma conditions and from line-to-continuum ratios to determine absolute abundances of elements from oxygen to iron. As will be discussed later, the RDS instrument is capable to measure emission lines and (uniquely) the X-ray continuum. A special arrangement of crystals facing each other on the opposite sites of the rotating drum will allow precise line Doppler shifts in a way similar to those incorporated in the Diogeness instrument \cite{Phillips2018}. The novelty of the approach incorporated in RDS to obtain the X-ray spectra is its possible very fast scanning - 10 drum turns per second. With such frequent rotations, the Bragg-reflected spectrum would normally be of low statistical quality (see Fig.~\ref{RDS_spectrum}), but it will be possible to built-up count statistics in time by assigning every recorded photon position to respective wavelength.

\begin{figure}
\centerline{\includegraphics[,width=0.6\textwidth,clip=]{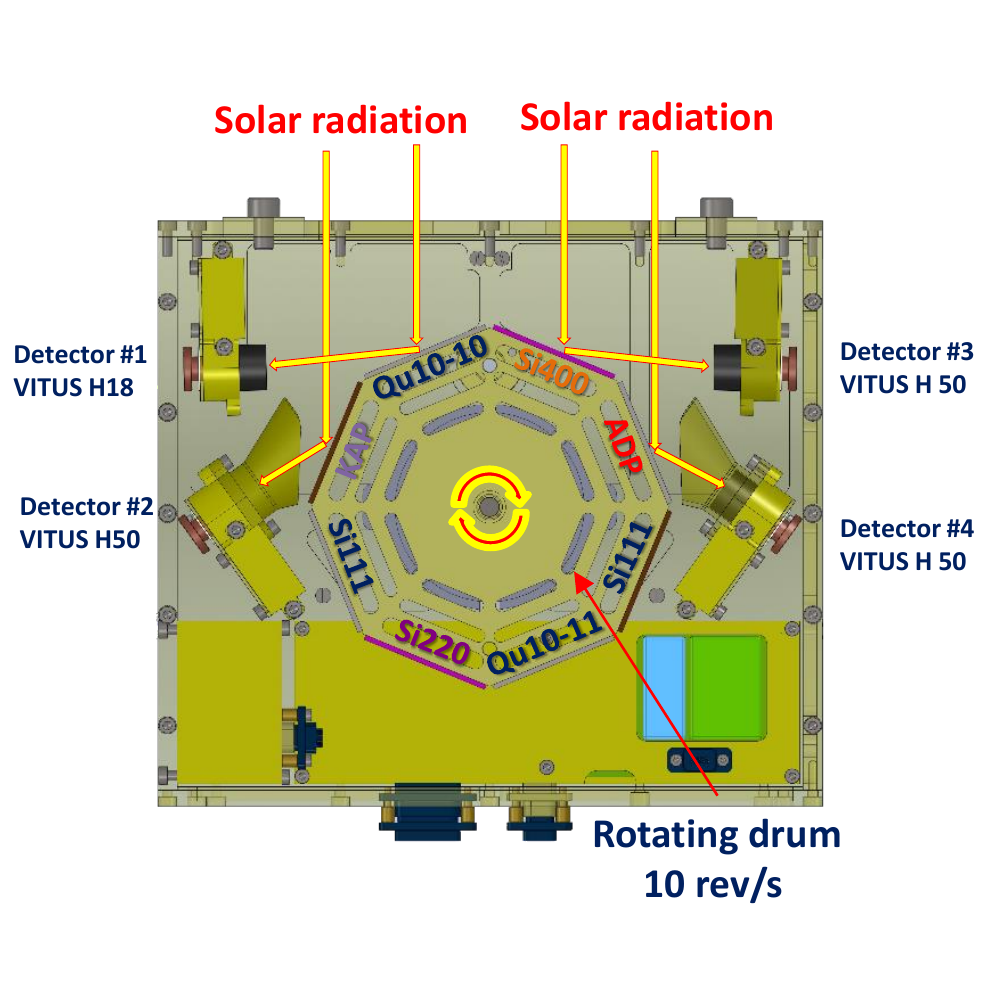}}
\caption{The RDS main components. Eight flat crystals (two of them identical: crystal cut Si 111) are fixed to the rotating (10\,rps) or rocking drum. Bragg reflected spectra are recorded by four SDD detectors. For every full rotation each of the eight crystals illuminate every of the four detectors over a short time interval.}
\label{fig_rds_inside}
\end{figure}

This will be done by using the exact timing information on the photon arrival time on the Silicon Drift Detectors (SDD). In RDS, we use four such state-of-the-art devices, in which the detection time can be established with an accuracy of $\sim$1\,$\mu$s. In one microsecond, the drum with all the crystals will rotate by $\sim$10\,arcsec only, so the line wavelength uncertainty corresponding to such a small time interval (respective change in the incidence angle) is less than the line width and therefore can be neglected. In relation with, the wavelength can be assigned to every photon detected with a high accuracy. This time-rotation angle-wavelength fixed relation will allow to build statistically significant spectra by summing photons arriving at the specified wavelength intervals in order to build enough statistics.

 \begin{table}[t]
\centering
\caption{\textsl{Characteristics of crystals used in RDS}}
\label{tab-crystals}
 \begin{tabular}{cccccccc}
  \hline
  No. & Crystal  & Orientation   & 2d & Wavelength range & Energy range  & Spectral & line FWHM  \\
  &&& [\r{A}]&[\r{A}]&[keV]&resolution [m\r{A} (eV)]&[m\r{A} (eV)]\\
   \hline
   \multicolumn{8}{c}{front detectors}\\
      \hline
  1 &  Si & 400& 2.715 & 1.678--2.330 &5.321--7.389 &  0.11 (0.336) &0.34 (1.234) \\
  2 &  Si & 220& 3.840 & 2.375--3.295 &3.763--5.220 & 0.15 (0.237) & 0.65 (0.806) \\
  3 &  $2\times$Si & 111& 6.2715 &3.878--5.381  &2.304--3.197 &0.25 (0.145) &0.97 (0.537) \\
  4 &  Quartz & 10-11&6.684 & 4.133--5.735 &2.162--3.000 &0.27 (0.136) & 0.98 (0.545) \\
  5 &  Quartz & 10-10& 8.514 &5.265--7.306  &1.697--2.355& 0.35 (0.107) & 1.3 (0.421)\\
  6 &  ADP & 101& 10.648 & 6.585--9.137  &1.357--1.883 &0.44 (0.085) & 1.3 (0.239)\\
  7 &  KAP & 001& 26.640 & 16.474--22.859  &0.542--0.753 &1.1 (0.034) & 6.5 (0.207)\\
     \hline
     \multicolumn{8}{c}{rear detectors}\\
        \hline
  1 &  Si & 400& 2.715 &0.442--1.618  &7.663--28.051 &0.15 (1.730)&0.39 (2.133)\\
  2 &  Si & 220& 3.840 & 0.625--2.288  &5.419--19.837&0.22 (1.221)&0.34 (1.238) \\
  3 &  $2\times$Si & 111& 6.2715 &1.021--3.737  &3.318--12.143 &0.36 (0.748) &0.34 (1.242) \\
  4 &  Quartz & 10-11&6.684 & 1.088--3.983  &3.113--11.396 &0.38 (0.702)& 0.35 (1.272)\\
  5 &  Quartz & 10-10& 8.514 & 1.386--5.073  &2.444--8.945 &0.49 (0.551) & 0.66 (0.821)\\
  6 &  ADP & 101& 10.648 & 1.734--6.345   &1.954--7.150 &0.61 (0.441)  & 0.67 (0.580)\\
  7 &  KAP & 001& 26.640 &4.337--15.875  &0.781--2.859 &1.5 (0.176) &1.9 (0.192)\\
  \hline
\end{tabular}
\end{table}

As shown in Fig.~\ref{fig_rds_inside}, the flat mono-crystals are attached to an octagonal drum rotating at high rate or in a slow rocking mode. The crystal mount geometry is selected such that for most of time, the solar X-rays illuminate a pair of the rotating crystals, from which the Bragg-reflected photons are recorded by two pairs of SDD detectors. In time, every detector records spectra at a slightly different Bragg angle. The respective crystal incidence angles can easily be determined and converted to corresponding incoming photon wavelengths (or equivalent energies) by using a calibrated rotation angle readout system. In the front panel of the instrument the two apertures are covered by corresponding energy assigned filters (see Fig.~\ref{fig_rds}b) that block solar thermal, optical and EUV radiation.

The symmetric location of detectors on both sides of the rotating drum (with respect to the direction towards the Sun) represents the configuration fulfilling tasks of a ``Dopplerometer'' \cite{Sylwester_etal2015}.

In the RDS we use eight flat crystals, two of them identical (see Table~\ref{tab-crystals} for details). For each crystal two spectral wavebands are given. The first band is observed by the ``front'' pair of detectors (those closer to the Sun). The second waveband is observed by the rear detectors' pair. Two identical crystals of Si~111 are placed in the ``classical'' Dopplerometer configuration. With all selected rotating crystals a wide continuous spectral range extending from 0.4\,\AA~to~22.8\,\AA\ will be covered in 0.1\,s. This will allow us to determine spectral line fluxes and Doppler shifts for many abundant elements (from oxygen to iron) emitted from flaring plasma with temperatures between 1\,MK and 50\,MK. These data will assist the measurements of spectra made using the B-POL unit.

The selection of detectors for RDS experiment was defined by scientific requirements such as desired wavelength range, suitable active area and the fast readout. Since the spatial resolution was not an option, the Si-PIN detector and SDD were considered. We have chosen the SDD because of its better performance as compared with the Si-PIN of the same area. An excellent energy resolution is achieved due to very low detector capacitance of SDDs. Although Bragg reflection is the primary technique for defining the photon energies, we use also direct energy measurement as is possible with SDD as complementary information. This will allow detecting higher orders of Bragg reflections or rejecting fluorescent/background counts. Detector energy resolution is important, although not critical in this respect.

Selected detectors for RDS experiment are listed in Table \ref{tab-detectors}. These modules produced by KETEK, GmbH contain SDD chip with the JFET, feedback capacitor, reset diode, thermistor and thermoelectric cooler (TEC) integrated in hermetic TO-8 package [https://www.ketek.net/]. Both modules H50 and H18LE have a chip thickness of 450\,$\mu$m and active area limited by internal multilayer collimator. Larger modules are also available, but with thicker Be windows. Selection of H50 is a trade-off between active area and Be window thickness which determine the low energy cutoff. The 12.5\,$\mu$m Be window has a transmission of $\sim$0.25 for 1\,keV photons. This gives a reasonable sensitivity for most of energy ranges listed in Table \ref{tab-crystals}. Energy ranges obtained from KAP crystal extend down to $\sim$0.5\,keV, thus a single H18LE module with polymer AP3.3 window is employed to cover the KAP energy ranges (see Fig.~\ref{window_tr}). Theoretical transmission curve for AP3.3 window has been calculated based on \cite{Scholze2005}.

\begin{figure}
\centerline{\includegraphics[,width=1.0\textwidth,clip=]{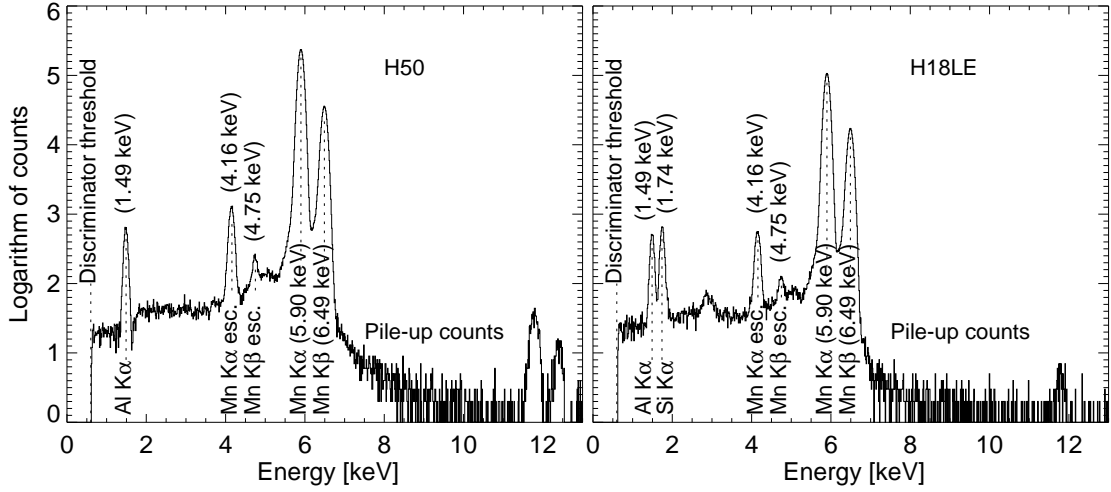}}
\caption{Histograms of energy spectrum taken in laboratory conditions using $^5$$^5$Fe source and selected detectors.}
\label{RDS_2}
\end{figure}

The results of laboratory tests of H50 and H18LE detectors are shown in Fig.~\ref{RDS_2}. The measurements were taken using $^5$$^5$Fe source placed at close distance of $\sim$2\,mm in front of detector window. Thin Al foil has been put between the source and the window to obtain additional Al K$\alpha$ line at 1.49 keV. As a readout system Digital Pulse Processor (DPP) with trapezoidal shaping was used. Detailed information on measurements are gathered in Table \ref{tab-measurements}. Energy threshold was set to $\sim$0.6\,keV. Both spectra contain clear Al K$\alpha$, Mn K$\alpha$ and K$\beta$ peaks as well as Mn escape peaks at 4.16 and 4.75 keV. An additional line at 1.74\,keV (Si K$\alpha$) for H18LE can also be seen, most likely due to fluorescence from the support grid of the entrance window. All counts above Mn K$\beta$ are due to pile-up with clear peaks at twice the Mn K$\alpha$ and sum of Mn K$\alpha$ and K$\beta$ energies. Intensities of pile-up counts are different for H50 and H18LE because of different count rates between the measurements. The lower count rate in case of H18LE was due to a smaller active area as well as lower overall transmission of the AP3.3 window. To compensate for the lower count rate, H18LE spectra integration time was twice as long in the example shown. The energy resolution obtained was 147\,eV FWHM at 5.9\,keV for both detectors.

\begin{figure}
\centerline{\includegraphics[,width=1.0\textwidth,clip=]{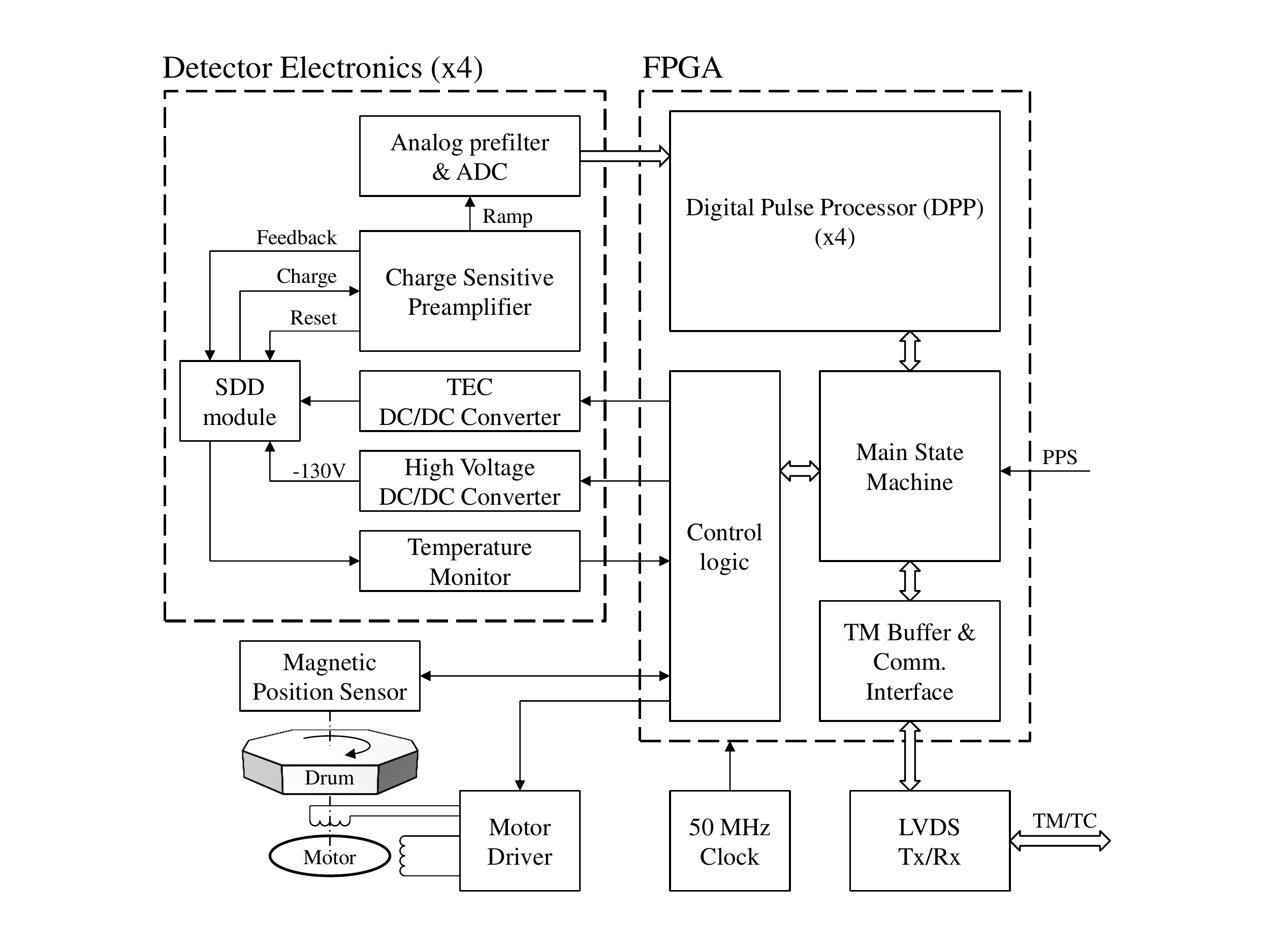}}
\caption{Block scheme of RDS electronics design.}
\label{RDS_3}
\end{figure}

\begin{figure}
\centerline{\includegraphics[,width=1.0\textwidth,clip=]{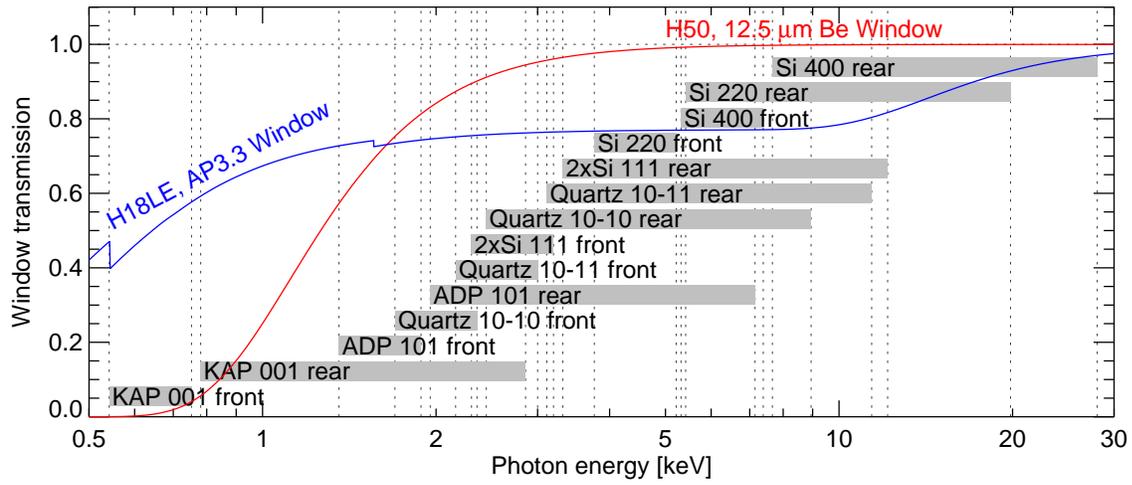}}
\caption{The RDS detectors window transmissions. The observed spectral energy intervals for respective crystals are marked in grey.
}
\label{window_tr}
\end{figure}

To operate the SDD, several subsystems are necessary as shown in Fig.~\ref{RDS_3}. Each detector is connected to the individual electronics block. SDD achieves the best performance when operated at temperatures below -20$^{\circ}$C, thus the built-in TEC will be used for active cooling. The TEC is supplied by a dedicated DC/DC converter with the power controlled by the FPGA logic. On the other hand, the detector mounts are connected to the KORTES radiator through active heat pipes (as shown in Fig.~\ref{fig_rds}a) to radiate the heat from the hot side of the TEC. Detector temperatures will be monitored by the thermistor mounted close to the SDD chip on the cool side of the TEC. Temperature monitor circuitry converts the thermistor resistance to digital values. The look-up table in FPGA provides the measured temperature on a centigrade scale. This allows for automatic close-loop temperature stabilization at the requested level.

The other DC/DC converter provides High Voltage (HV) of -130 V. This voltage is required to bias the SDD ring electrodes and back contact through a voltage divider and filtering circuitry.

The Charge Sensitive Preamplifier (CSP) converts a charge induced due to photo-electric effect into voltage pulse. The CSP is a ``reset type'', thus the output is a ramp signal consisting of small steps of several mV. Each step corresponds to interaction of X-ray photon with the detector. The step amplitude is proportional to the energy deposited by the absorbed photon. CSP uses SDD integrated JFET as a first stage of the amplifier and feedback capacitor. CSP contains also dedicated reset circuitry to discharge the feedback capacitor (output ramp reset).

The ramp output is fed to signal conditioning circuitry with a high pass filter converting the voltage steps to exponential pulses. Such a signal is then digitized by 12-bit ADC at sampling rate of 50\,MS/s. Next the detector signal is processed by DPP implemented in FPGA to extract desired information. Unlike the commercially available DPPs which offer great flexibility, the RDS design is unique and instrument-specific. The functionality is tailored for RDS requirements and the parameters such as peaking time, energy range and others are fixed while a stronger effort was taken to improve the system reliability and stability.

The information extracted from the detector signal are photon arrival time and energy. The arrival time is a key data while the energy is an auxiliary one. FPGA core contains a precise time counter with 1\,$\mu$s step. Each detected photon is put to telemetry buffer as a separate data packet with a time stamp taken from the microsecond counter. The same time stamps are assigned to other data packets with current drum position measured once every $\sim$96\,$\mu$s time interval. In this way one can determine an exact drum position at the moment of photon registration. Photon energy from direct detector measurement is used for event validation (which must be consistent with expected energy due to Bragg reflection at a given drum position). This method allows for detecting higher orders of Bragg reflection, fluorescent counts from the crystals or estimating background counts between effective drum position ranges.

RDS does not use a microprocessor, neither separate IC nor FPGA soft processor. Since the IDPU (inside B-POL) is common for all SolpeX blocks, RDS has only dedicated state machines implemented in FPGA to handle telemetry packets generation, command execution and communication with IDPU through a serial link using Low Voltage Differential Signaling (LVDS) physical interface.

\begin{table}[t]
\centering
\caption{\textsl{RDS Detectors}}
\label{tab-detectors}
 \begin{tabular}{cccccc}
  \hline
  No. & Detector  & Location   & Active area [mm$^2$]  &Window &Package  \\
   \hline
  1 &  KETEK VITUS SDD H18LE & front &18& AP3.3 polymer & filled with Xe\\
  2 &  KETEK VITUS SDD H50 & rear &50& 12.5\,$\mu m$ Be &evacuated \\
  3 &  KETEK VITUS SDD H50 & front  &50& 12.5\,$\mu m$ Be&evacuated \\
  4 &  KETEK VITUS SDD H50 & rear &50& 12.5\,$\mu m$ Be &evacuated\\
  \hline
\end{tabular}
\end{table}

\begin{table}
\centering
\caption{\textsl{Results of RDS detectors' tests}}
\label{tab-measurements}
 \begin{tabular}{cccccccc}
  \hline
  No. & Detector  & Peaking  & Det.  & Duration & Av. count  & Total Histogram  & Energy       \\
  &&time&Temp.&&rate&counts&resolution\\
   \hline
  1 &  H50        & 1.2\,$\mu$s  & -35$^{\circ}$C & 0.5 h    & 1937 c/s       &  3 445 892             & 147 eV FWHM at 5.9 keV \\
  2 &  H18LE      & 1.2\,$\mu$s  & -25$^{\circ}$C & 1 h      & 436 c/s        &  1 569 016             & 147 eV FWHM at 5.9 keV \\
  \hline
\end{tabular}
\end{table}

\begin{figure}
\centerline{\includegraphics[,width=1.0\textwidth,clip=]{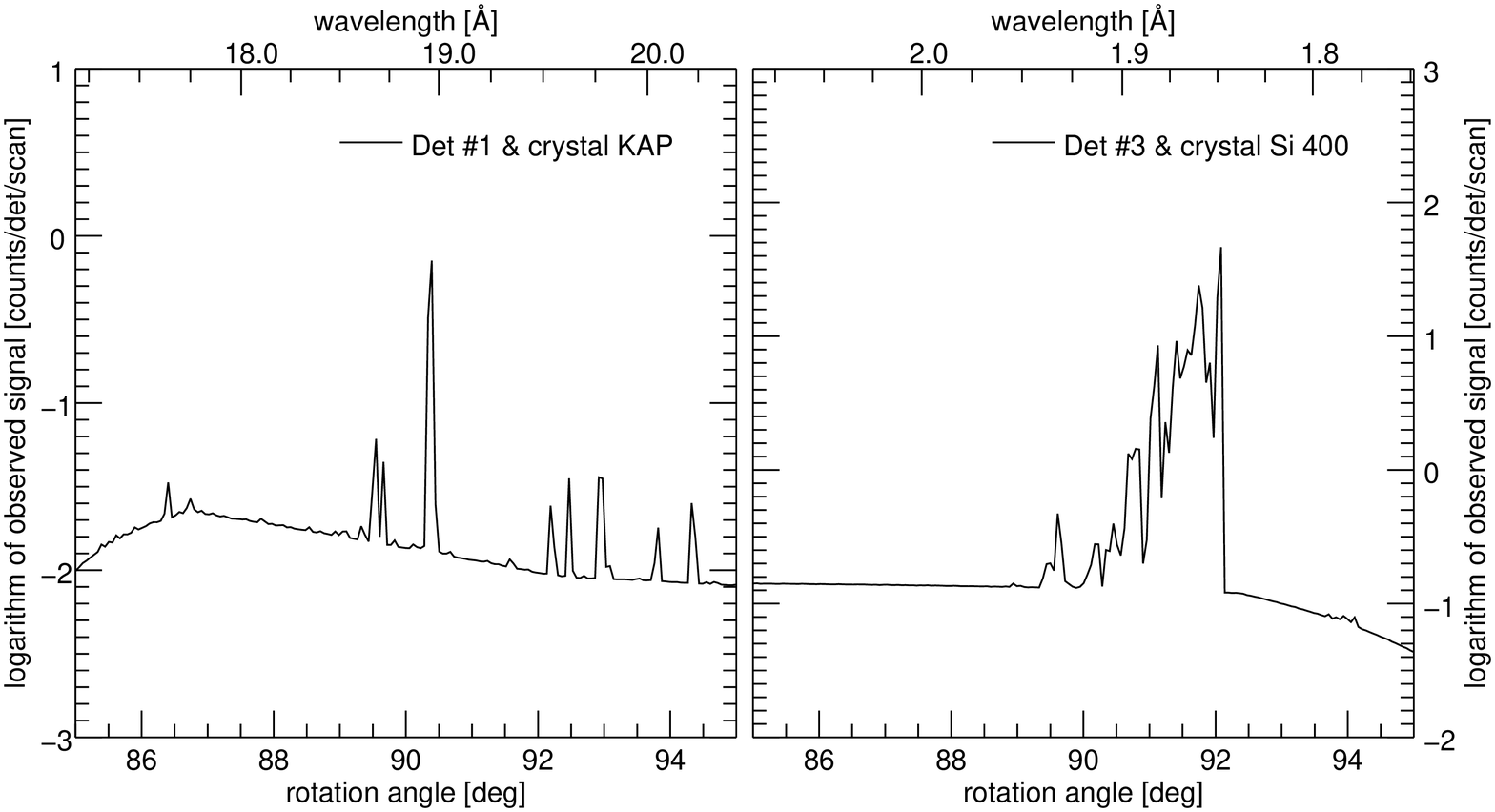}}
\caption{Example of RDS spectra to be observed in the rocking mode, assuming that the scan consists of 178 discrete drum angles and photons were recorded at each of them for 1\,ms. In the spectral simulations, we assumed that the source (flare) is of M5 {\em GOES} class.}
\label{RDS_scan}
\end{figure}

The rotating drum is actuated by a stepper motor with 200 steps per revolution. Moreover, the motor driver allows for microstepping control with 32 microsteps/step. The RDS can operate in two modes:
\begin{itemize}
\item Rotating mode (default) - the drum is rotated continuously (left or right direction). Although the motor speed can be adjusted in wide range, the fixed speed of 10\,RPS has been chosen as optimal in terms of motor efficiency and rotation stability. Thanks to a moment of inertia of the rotating drum, the torque ripple is reduced to marginal level and thus the rotation angular velocity is considered as constant. Since the full revolution lasts 0.1\,s, this is the basic time interval for spectra integration. However, even for strong flares, much longer time interval is required to collect a good count statistics.
\item Rocking mode (on demand) - the drum is rotated forth and back within specified range of angles for narrow-wavelength band scans. The motor is driven in microstepping mode yielding a 6400 discrete drum angles per revolution (i.e. 202.5\,arcsec/microstep). This gives, for instance, $\sim$178 discrete drum positions within a range of 10 degrees (see examples shown in Fig.~\ref{RDS_scan}). The maximum angular velocity in rocking mode is limited to 1\,microstep/1\,ms (i.e. $\sim$0.156\,RPS). This is due to requirement of collecting at least 10 drum position measurements for each microstep in order to improve the S/N ratio on the drum position estimate.
\end{itemize}

The drum position is measured by a magnetic sensor with a resolution of 12-bits (i.e. 4096 values) per revolution. The quantization error and the measurement noise requires a reconstruction of the drum position by fitting a linear function to a time series of data (this will be done
by ground software). In order to estimate expected errors of the reconstructed drum positions, we carried out a dedicated analysis. In case of rotating mode, fitting the linear function to 1\,s series of data ($\sim$10000 drum position's measurements), results in accuracy better than $\pm$20\,arcsec,
which has no impact on the spectral resolution. Reconstruction of the drum position from longer series of data can improve the accuracy even more. In rocking mode, the function of drum angle in time is more staircase-like than linear due to discrete drum positions. Moreover, the data series for linear fitting is limited to length of single scan (typically $\sim$1800 drum position's measurements within 10\,degrees scan). This results in lower accuracy of reconstructed drum position. The error is in the range $\pm$140\,arcsec, which affects the spectral resolution. On the other hand, a great advantage of rocking mode is much better statistics since there is no dead drum positions as present in the rotating mode.

\begin{table}[t]
\centering
\caption{\textsl{Rocking mode scans}}
\label{tab-scans}
 \begin{tabular}{cccccc}
  \hline
  No. & Detector \#1 & Detector \#2  & Detector \#3 &Detector \#4 & Type of solar activity \\
   \hline
  1 &  O~{\sc xvii}&&Fe~{\sc xxv} &&quiet Sun, AR, flare\\
  2 &  O~{\sc xvii}&Fe~{\sc xxv}& &Fe~{\sc xxv}&flare\\
  3 &  &Ca~{\sc xix} \& Ar~{\sc xvii}& &Mg~{\sc xi}& AR, flare\\
  4 &  Si~{\sc xiii}&&Si~{\sc xiii} && AR, flare\\
  5 &S~{\sc xiv}  &&Ar~{\sc xvii} && AR, flare\\
  6 &Si~{\sc xiii}  &&S~{\sc xv} &&quiet Sun, AR, flare\\
  7 &  &&Ca~{\sc xix} && flare\\
  \hline
\end{tabular}
\end{table}

RDS will operate by default in the rotating mode for most of the time. Rocking mode will be activated occasionally upon command. In order to plan the operation of RDS abord {\em ISS}, we considered the following criteria:
\begin{itemize}
\item assumed that in the 2021--2023 observation period, flaring activity will be similar to that observed by {\em GOES} in the analogous period during the 24th activity cycle (years 2011--2012);
\item the fact that the uninterrupted time interval of seeing the Sun from {\em Nauka ISS} pointed platform is $\sim$12\,min/orbit;
\item the frequency of stronger flares is decreasing with increasing X-ray class of the event;
\item the duration of rise and decay phases of flares increases with flare class.
\end{itemize}

\begin{figure}
\centerline{\includegraphics[,width=1.0\textwidth,clip=]{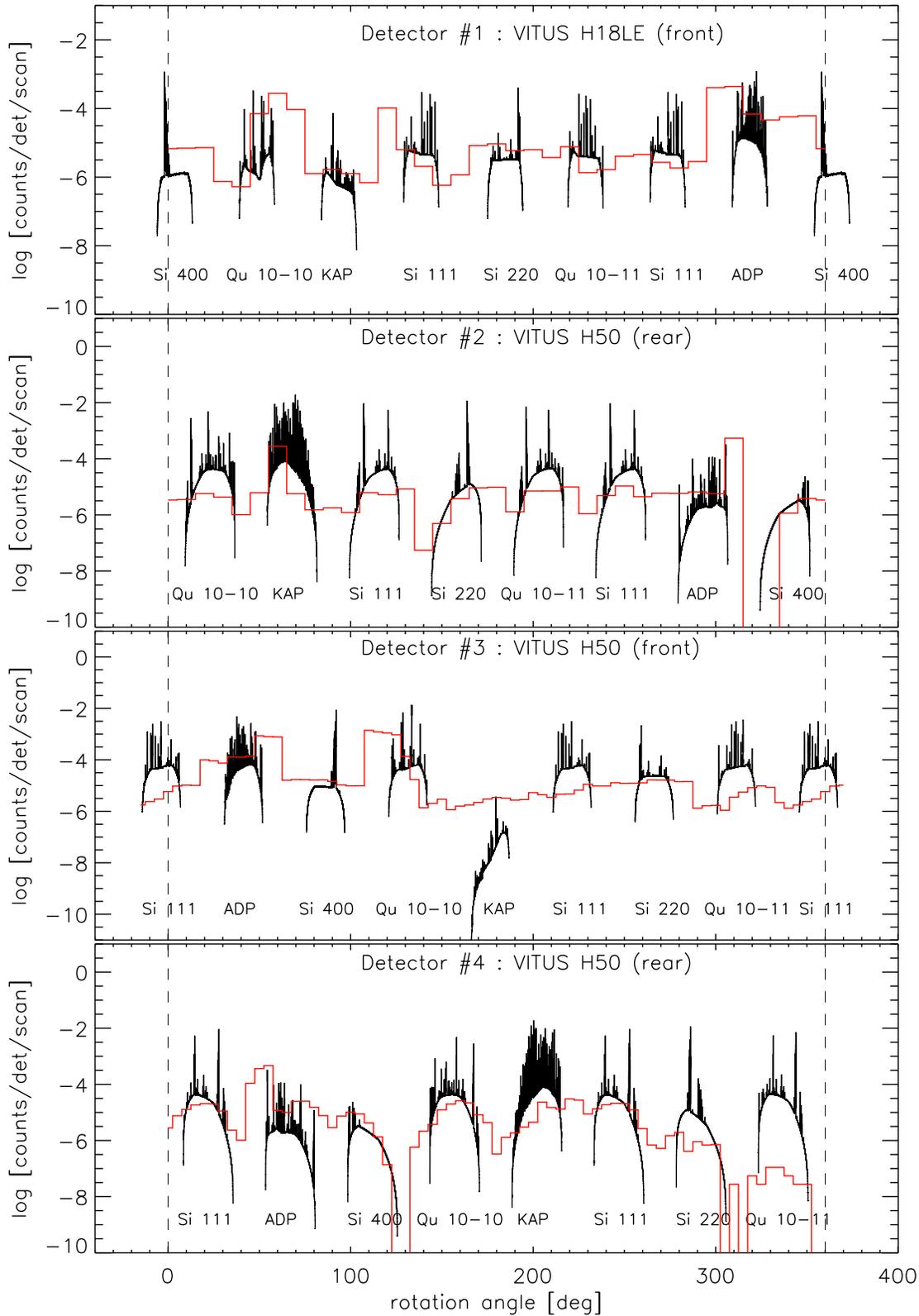}}
\caption{Expected spectral records to be seen by the four RDS SDD detectors in rotating mode. On the horizontal axis the rotation angle is plotted. On the vertical axis, the log of the count rate per one drum revolution is plotted. The point-like source characteristics are representative for an M5 class flare (in  {\em GOES} classification). The rotation rate is assumed to be 10\,revolutions/s. In order to obtain good statistics, count rates from many rotations should be collected in the case of weaker events. The red histogram levels correspond to expected background fluorescence. For the most of channels this level is below the ``real'' X-ray continuum.}
\label{RDS_spectrum}
\end{figure}

Using properties of all elements on the optical path of the individual SolpeX blocks (such as crystal reflectivity profile, filter transmissions and sensitivity of detectors), the signal to be collected over a single data gather interval DGI can be calculated according to the formula:
\begin{equation}
N[cts]=\frac{F E A}{\omega}
\label{referencja}
\end{equation}
here $A$ is the detector area (in [cm$^2$]), $\omega$ is the scanning angular velocity and $E$ is the efficiency function, that contain the instrument filters' transmission function and the detector efficiency. $F$ is given by:
\begin{equation}
F=\int\limits_{0}^{\infty} f(\lambda)R(\lambda)d\lambda
\end{equation}
\noindent where $f(\lambda)$ is the flux from an on-axis point source (in [phot cm$^{-2}$ s$^{-1}$]) and $R(\lambda)$ is the crystal reflectivity profile (rocking curve). In order to calculate the spectral signal to be collected by a detector, we assumed that the source is an M5 {\em GOES} class flare (T=17.22\,MK, EM=2$\times$10$^{49}\,$cm$^{-3}$). We calculated the expected spectral X-ray irradiance of the flare using CHIANTI \cite{DelZanna2015}.

An example of spectra observed on each detector during a single rotation of the drum are presented in Fig.~\ref{RDS_spectrum}. The red line represents the calculated fluorescence level. Fluorescence calculations were performed using Geant4 CERN package as described in \cite{Makowski2018}.
The observed histogram of spectra will be built as time progresses, with the count statistics depending on the source intensity. For flares above M5.0 class, enough photons will be recorded to study spectra variability on a time scale of $\sim$1\,s.

\begin{figure}
\centerline{\includegraphics[,width=1.0\textwidth,clip=]{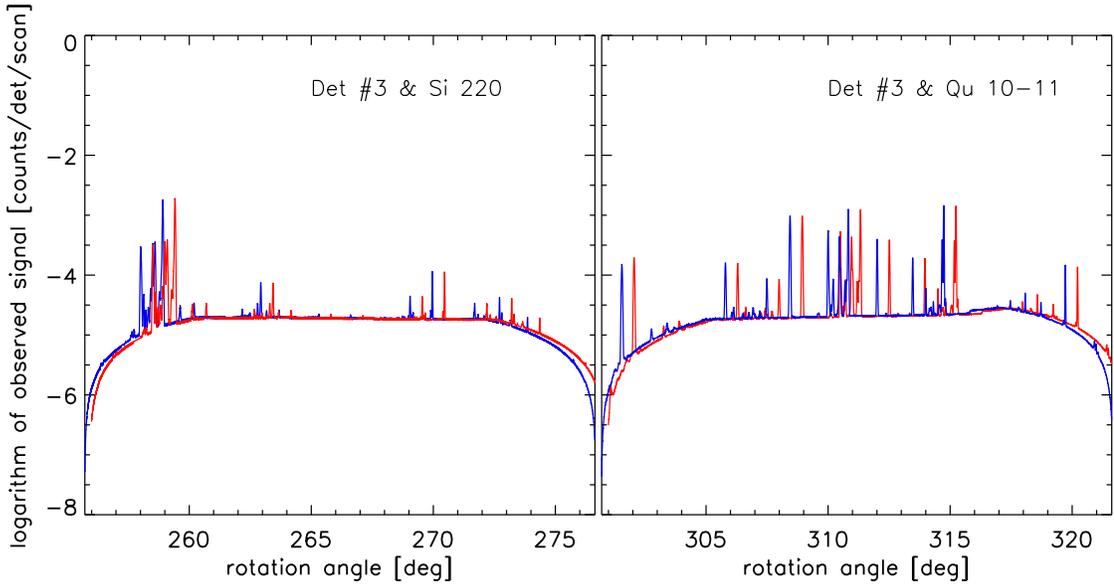}}
\caption{Example of RDS observations when two identical {\em GOES} M5 flares occur simultaneously on the Sun. The spectra marked with the blue and red lines come from flares located at two opposite edges of the solar disk respectively.}
\label{2sources}
\end{figure}

Rarely, it may happen the situation where there will be two or more active flaring regions (see discussion in Section \ref{section:phi}). An example of a spectrum observed by RDS when two {\em GOES} M5 class flares occur on opposite limbs of the solar disc is shown in Fig.~\ref{2sources}.

In the rocking mode it will be possible to observe fast variations of spectra in the limited, pre-defined spectral ranges covering interesting spectral bands. These bands of interest will be defined based on the inspection of spectra obtained in the long lasting fast rotation mode, although they will certainly include the triplet lines of O~{\sc vii}, useful in determining the plasma densities \cite{Wolfson1983}, \cite{Linford1988}, Fe~{\sc xvii} and Ne~{\sc ix} lines and other triplets of Mg~{\sc xi}, Si~{\sc xiii}, Ar~{\sc xvii}, Ca~{\sc xix} and Fe~{\sc xxv}, depending on activity level. Also the components of the Mg~{\sc xii} Ly$\alpha$ and O~{\sc viii} spin doublets will be studied. Such narrow-range scans will be especially useful when investigating initial flare phases for the presence of Doppler-shifted line components (Table \ref{tab-scans}).


\section{The B-POL X-ray polarimeter --- B-POL}
\label{section:bpol}
\begin{figure}
\centerline{\includegraphics[,width=1.0\textwidth,clip=]{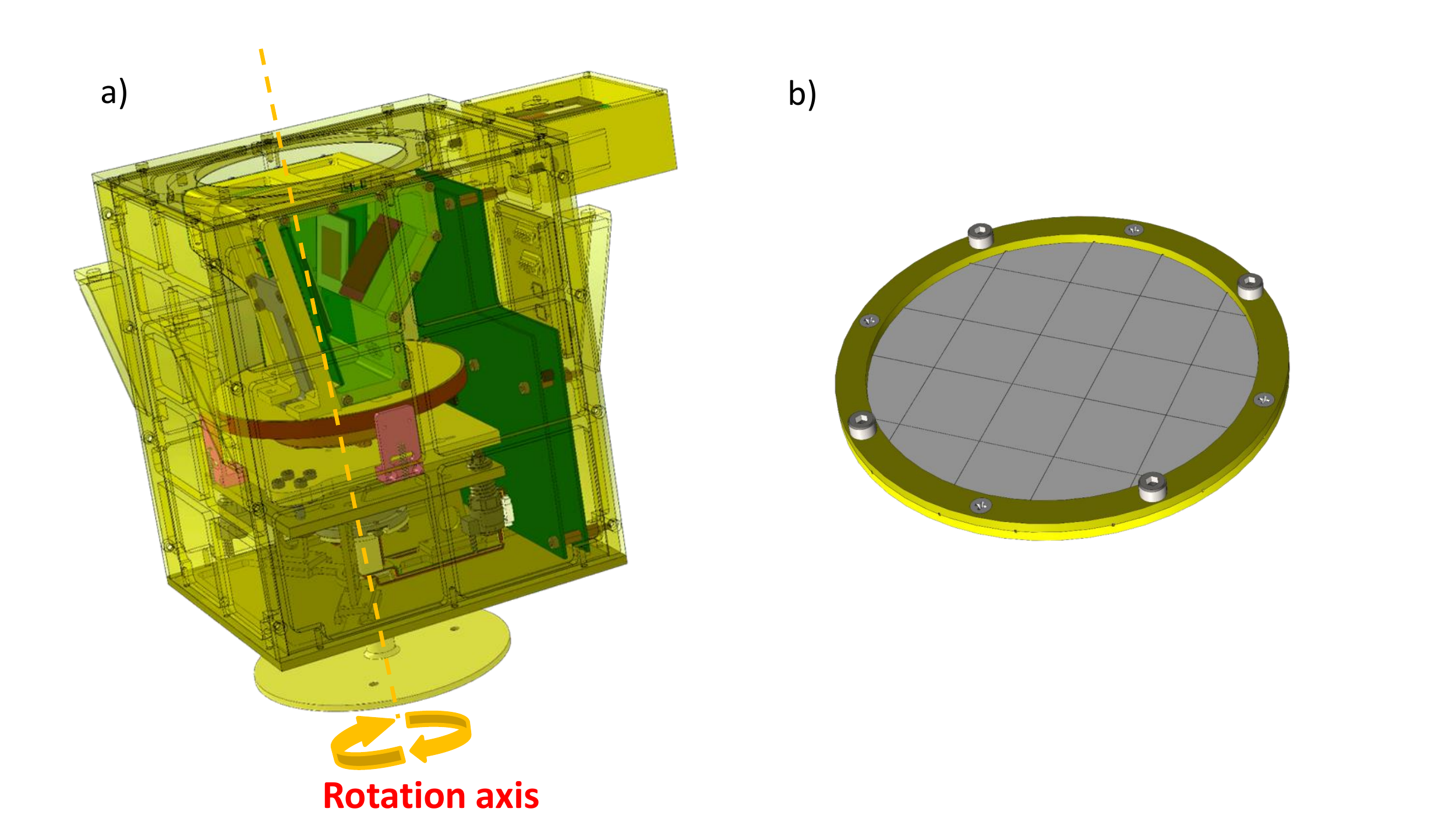}}
\caption{{\bf a)} Construction of B-POL measurements block from a CAD drawing. The rotation axis pointing towards the flare is indicated. {\bf b)} B-POL filter (placed on the front panel of KORTES). This filter prevents UV, visible, and IR radiation from reaching to the inside of the instrument, while not overly attenuating X-rays in the range from 3.9 to 4.1\,\AA. The transmission of the filter in this narrow spectral range is approximately $\sim$85$\%$. The diameter of the filter is 104\,mm. The chosen window material is 10\,$\mu$m of polyimide coated on both sides with 1200\,\AA\,of aluminium. The polyimide window is supported by Kevlar mesh (in black) where the thickness of the threads is 0.60\,microns.}
\label{Bpol_1}
\end{figure}

\begin{figure}
\centerline{\includegraphics[,width=1\textwidth,clip=]{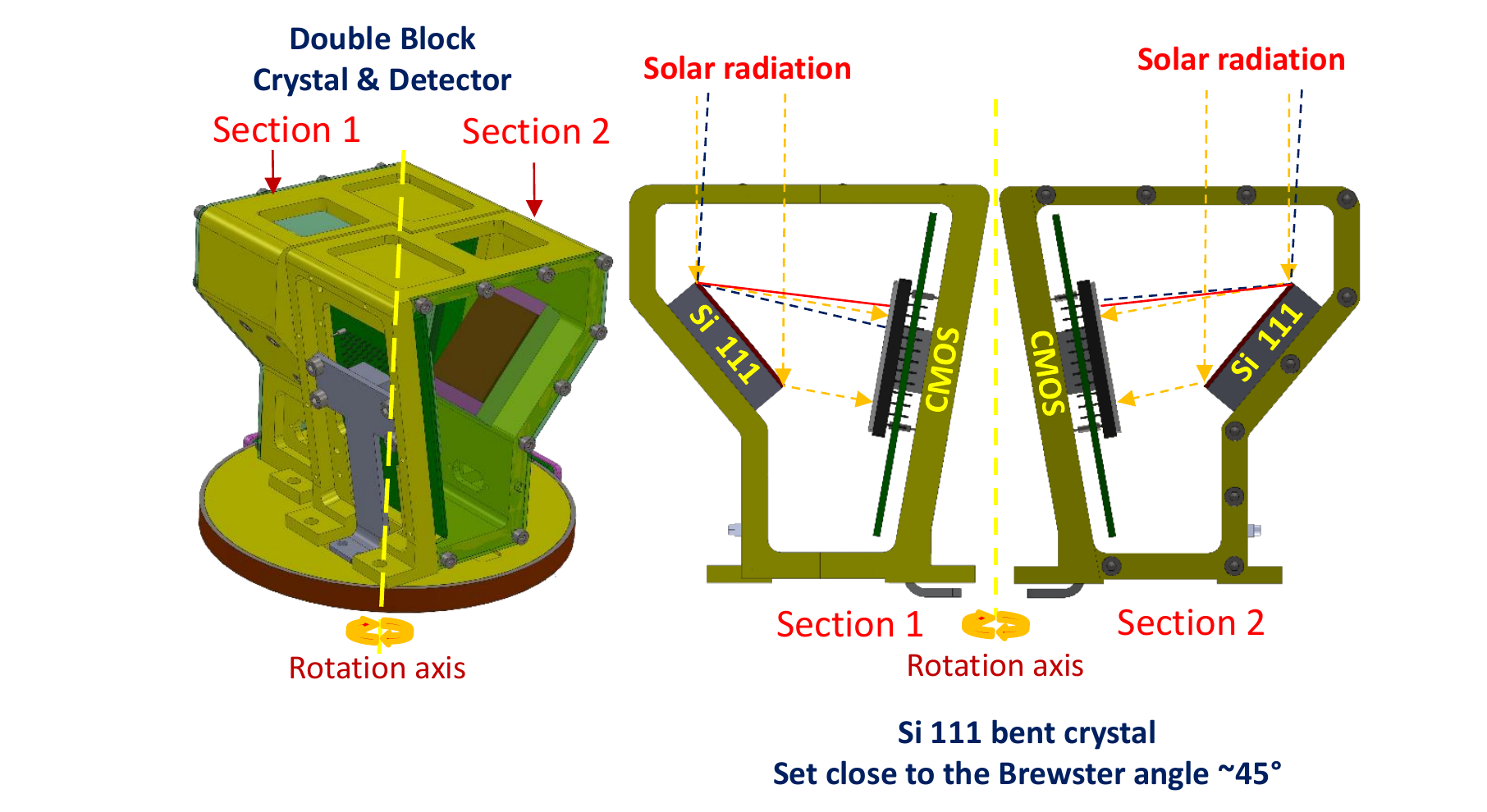}}
\caption{{\bf a)} The polarimeter of B-POL; The rotation axis (1 rotation/second) is to be pointed towards the flare. {\bf b)} A scheme of reflections from Si crystal bent to radius 821\,mm. The orange line in b) represents ray path in case the source is on the direction of rotation axis. In case it is somewhat off (blue rays), the spectral line positions will move back and forward as the table rotates, allowing thus to estimate the misalignment angle and automatically correct the rotation axis orientation.}
\label{Bpol_inside}
\end{figure}

The concept of B-POL X-ray polarimeter is based on the property that for Bragg reflections close to Brewster angle ($\sim$45$^\circ$), the reflection efficiency for illuminating linearly polarized radiation strongly depends on the angle between vector of the polarization direction and the vector normal to the crystal plane. If the angle is close to 45$^\circ$, the efficiency is maximized. In order to estimate polarization degree and orientation of the polarization plane, we will spin the crystal (i.e rotate the normal to the crystal plane) along the direction towards the source (flaring compact plasma volume in the corona), as shown in Fig.~\ref{Bpol_1}.

The B-POL unit is placed in the rear section of KORTES in order to keep the rotating radiator (fixed to the rotating table) in constant shadow (see Fig.~\ref{fig_kortes}a). In order to prevent the thermal/optical/XUV radiation entering KORTES, a special round thin filter is placed on the respective aperture in front of KORTES (Fig.~\ref{Bpol_1}b).

The heart of B-POL polarimeter (see Fig.~\ref{Bpol_inside}) consists of two identical, cylindrically bent mono-crystal wafers and two receiving large CMOS detectors (GSENSE400 BSI) fixed together at an angle of $\sim$45$^\circ$. The detectors have 2048$\times$2048 pixels, with the minimum read-out time of 48 frames per second. Both crystal-detector sections are fixed firmly to the table continuously rotating along the axis directed to the source. The source selection is based on the PHI X-ray disk image. The rotation rate of the table has been selected to be one revolution per second which represents a compromise between the CMOS read-out time (1/48\,s) and expected rate of polarization fluctuations (typically from seconds to minutes). The pointing of the rotation axis will usually be tracking the location of the brightest region of interest by inclining the support of rotating table to the desired direction within the angular range of $\pm$1$^{\circ}$ with an accuracy better than 1~arcmin. We selected to use Silicon~111 ideal mono-crystal wafer bent cylindrically to a 820.97\,mm radius as the dispersive medium. The adopted crystal-detector geometry set-up presented in Fig.~\ref{fig_bpol_spectrum}a allows spectra to be obtained in the spectral range 3.9\,\AA\,--\,4.1\,\AA\ with an exceptionally good spectral resolution of $\sim$90$\,\mu$\AA/bin (see Fig.~\ref{fig_bpol_spectrum}b). This resolution is better than the instrumental spectral resolution of $\sim$340$\,\mu$\AA\ which in turn is much less than characteristic thermal/turbulent width of emission lines. The crystal-detector unit geometry was evaluated using a method developed first for ChemiX instrument decribed in details in Appendix A of \cite{Siarkowski2016}.

There are two identical crystal-detector sections (see Fig.~\ref{Bpol_inside}) facing opposite directions mounted on the rotating table with dispersion planes co-aligned. Such a construction (called X-ray Dopplerometer arrangement \cite{Sylwester_etal2015}) will allow the precise determination of possible Doppler shifts of X-ray lines and check whether the rotation axis is indeed directed exactly towards the centre of flaring region. Using two identical independent co-rotating sections will allow the rotational modulation of the signal due to polarization to be separated from this caused by the source offset.

\begin{figure}
\centerline{\includegraphics[,width=0.8\textwidth,clip=]{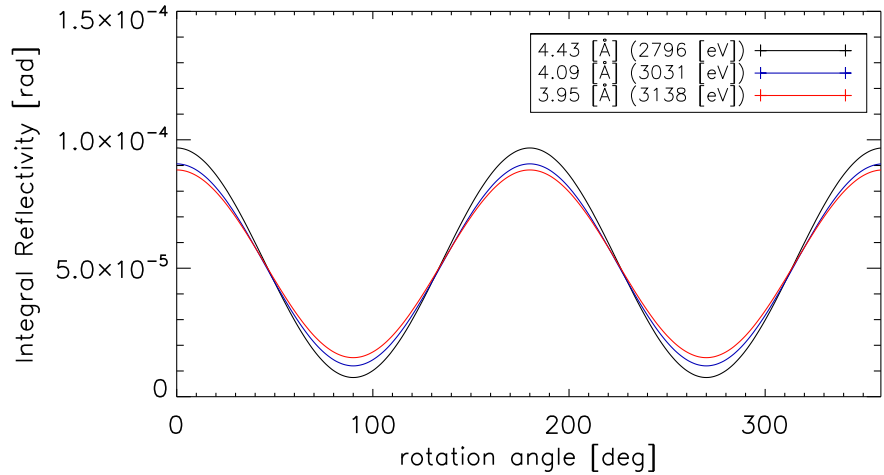}}
\caption{Changes of the crystal integral reflectivity during the rotation. We assumed that the X-ray source emitted radiation made of 80\% linearly polarised component and 20\% unpolarised component. The calculations were performed for the energy corresponded to the Brewster angle- 4.43\,\AA, and for the energies corresponding to two wavelengths at the edges of the observed spectral range. }
\label{polarization}
\end{figure}

During the rotation of the crystal-detector unit, the position of the crystal dispersion plane with respect to incoming light beam will continuously change. This will give rise to a modulation of the reflected beam intensity provided the incident X-ray beam is linearly polarized. It has been shown with the Electron Beam Ion Trap \cite{Beiersdorfer_etal1996} that the polarization of the He-like triplet lines for iron and lighter elements can be substantial and depends on the relative orientation of the EBIT particle beam and the crystal polarimeter plane. From Fig.~1 of \cite{Beiersdorfer_etal1996} it follows that, for Ar~{\sc xvii} (as well as for much weaker Cl~{\sc xvi} triplet lines), the degree of polarization of \emph{w} and \emph{x} lines can be close to 60\% and -60\% respectively. This level of polarization would easily be detected by B-POL for stronger flares. For incoming radiation being 100\% polarized, the depth of signal modulation will be also 100\% at the Brewster angle. For unpolarized radiation, no rotational modulation is expected. Fig. \ref{polarization} presents the modulation pattern of the integral reflectivity of the Si~111 crystal, calculated using the XOP package (\cite{rio11}). The integral reflectivity is one of the crystal characteristics, which directly affects the intensity of the spectra observed.

Using properties of filters and respective geometrical characteristics of crystal-detector units, the signal to be collected over a single data gather interval (DGI) can be calculated according to the formula introduced in \cite{Siarkowski2016}. The example of the spectrum to be measured for the M5.0 flare is given in Fig.~\ref{fig_bpol_spectrum}b.

\begin{figure}
\centerline{\includegraphics[,width=1.0\textwidth,clip=]{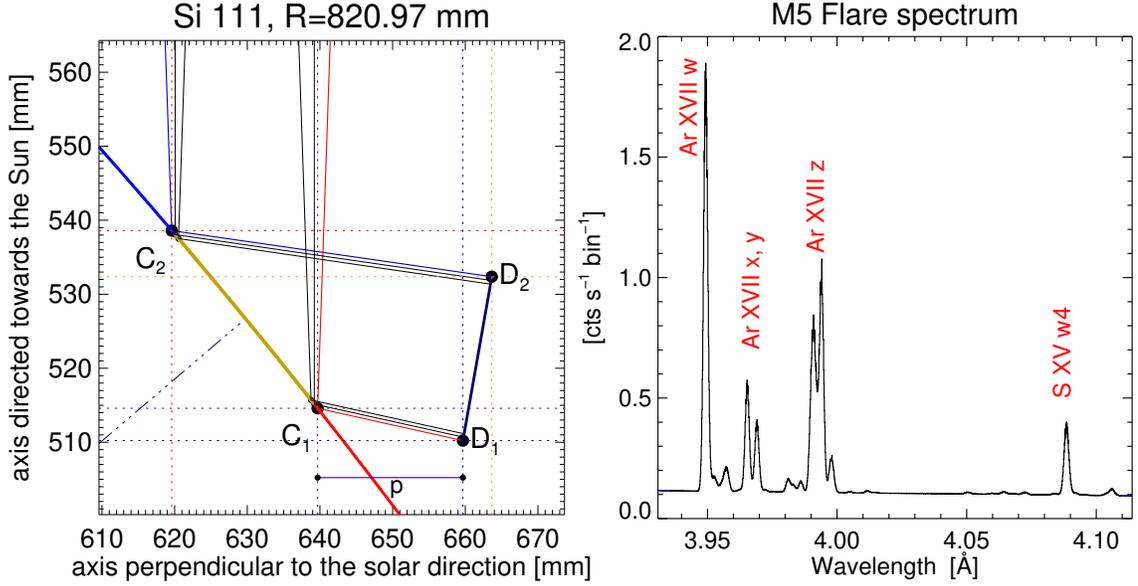}}
\caption{{\bf a)} Crystal-detector geometry of B-POL. Dots C1 and C2 indicate location of the crystal edges, D1 and D2 the detector. ``p'' is the projected distance from the lower edge of the crystal to the detector (p=20\,mm). {\bf b)} Corresponding M5 class flare spectrum to be recorded during one rotation of the crystal-detector section of B-POL is presented as calculated based on CHIANTI.}
\label{fig_bpol_spectrum}
\end{figure}

As indicated in Fig.~\ref{fig_bpol_spectrum}b, in the selected spectral range a number of strong emission lines are present, due to Ar~{\sc xvii}, Ar~{\sc xvi} and S~{\sc xvi}. The observed spectral range will be recorded for Bragg incidence angles close to the Brewster angle, and so their reflected intensities will depend on the degree of linear polarization and respective orientation of the polarization vector relative to normal of crystal surface.

The Ar as well as S lines are nearly always emitted by coronal sources activity above {\em GOES} B5.0 class level \cite{Sylwester_etal2010}. Therefore, the B-POL will record spectra not only for flares in progress but also for stronger active regions for most of the time. This will allow for studies of thermal plasma properties like turbulence, directed motions, differential emission measure distribution and elemental composition for Ar and S. However, lines of only few elements are present in the B-POL spectral range which was selected to optimize polarization detections. This high resolution spectral information will supplement that received from RDS.


\section{Summary}
The SolpeX spectrometer consisting of three functionally dependent blocks will be placed on the {\em ISS} as a part of KORTES Russian instrument. SolpeX consists of a simple pin-hole imager which will identify targets (bright X-ray sources in the solar corona) for the B-POL Bragg polarimeter. The RDS rotating spectrometer with eight flat crystals will provide measurements of X-ray spectra in a contiguous, wide wavelength range (1.3--23\,\AA) and will achieve maximum time resolution of $\sim$0.1\,s. With these three X-ray measurement units, the following observations will be performed for flares and/or active regions containing plasma of temperature above 1\,MK:

\begin{itemize}
\item individual AR/flare light curves in selected energy bands and their heliocentric coordinates with time resolution 0.1--1\,s,
\item flare linear polarization vector characteristics at energy $\sim$3\,keV every 1--10\,s,
\item AR/flare spectra in the entire soft X-ray range, with time resolution of a few seconds.
\end{itemize}

These measurements will complement each other allowing substantial progress in understanding the processes of magnetic energy release, transport and dissipation in various solar coronal structures, flares in particular. The success of this project will pave the way for building a larger instrument, based on similar principles, to be placed on dedicated future solar satellites.

At the time of writing, the instrument has passed initial design review and all system requirements, mechanical and electrical interfaces are fully defined and fixed. Complete mechanical construction of SolpeX has been manufactured. The B-POL unit including filter mounts underwent vibrational tests in May 2018. No problems were detected in the frequency range expected during the launch on Progress spacecraft. Most of electronics breadboards have been manufactured and tested at subsystem level. The instrument is currently at a phase of engineering model manufacturing and integration. Simultaneously, the FPGA cores are under development.

\begin{acknowledgements}
CHIANTI is a collaborative project involving George Mason University, the University of Michigan (USA) and the University of Cambridge (UK). We would like to thank Professor Ken J.H. Phillips for corrections and helpful comments. We acknowledge financial support from the Polish \emph{National Science Centre} grant 2013/11/B/ST9/00234. This research was partially supported by the Russian Science Foundation (project No. 17-12-01567).
\end{acknowledgements}

\bibliographystyle{spmpsci}      
\bibliography{Bib_file}

\end{document}